\documentclass{PoS}

\newcommand{\cV}{\mathcal{V}}
\newcommand{\cH}{\mathcal{H}}

\newcommand{\beqa}{\begin{eqnarray}}
\newcommand{\eeqa}{\end{eqnarray}}
\newcommand{\beq}{\begin{equation}}
\newcommand{\eeq}{\end{equation}}

\title{Aspects of confinement from QCD correlation functions }

\ShortTitle{Aspects of confinement from QCD correlation functions }

\author{\speaker{Christian S. Fischer}%
         \\
 Institute for Nuclear Physics, 
 Darmstadt University of Technology,\\ 
 Schlossgartenstra{\ss}e 9, 64289 Darmstadt, Germany\\
 GSI Helmholtzzentrum f\"ur Schwerionenforschung GmbH,\\ 
 Planckstr. 1,  64291 Darmstadt, Germany.\\
        E-mail: \email{christian.fischer@physik.tu-darmstadt.de}}

\author{Axel Maas\\
        Institut f\"ur Physik, Karl-Franzens Universit\"at Graz, 
Universit\"atsplatz 5, A-8010 Graz, Austria.\\
        E-mail: \email{axelmaas@web.de}}

\author{Jan M.~Pawlowski\\
        Institut f\"ur Theoretische Physik, University of
  Heidelberg,\\ Philosophenweg 16, D-62910 Heidelberg, Germany.\\
        E-mail: \email{J.Pawlowski@thphys.uni-heidelberg.de}}

\abstract{We discuss the properties of ghost and gluon propagators 
in Landau gauge Yang-Mills theory and their relation to the confinement
problem. In general two types of infrared behavior of these functions
are allowed from their functional equations: scaling and decoupling.
Both solutions show positivity violations in the gluon propagator and lead
to a confining Polyakov loop potential. However, only the scaling solution
agrees with the Kugo-Ojima confinement scenario and the related formulation 
of a physical Hilbert space of Yang-Mills theory. Our numerical results for 
the gluon dressing function agree almost pointwise with the lattice results
at all physical momenta.}

\FullConference{8th Conference Quark Confinement and the Hadron Spectrum\\
                 September 1-6, 2008\\
                 Mainz. Germany}

\begin{document}

\section{Global symmetries, confinement and the infrared behavior of Yang-Mills theory}

In this talk we are concerned with the infrared behavior of the 
dressing functions of the ghost and gluon propagators of QCD. There
has been much debate in the past years about the zero momentum limit
of these functions mainly due to an apparent mismatch between solutions
obtained from lattice gauge theory \cite{Cucchieri:2008fc,lvs} and functional 
equations in the continuum, i.e. Dyson-Schwinger equations 
\cite{Alkofer:2000wg,von Smekal:1997vx,Watson:2001yv,Lerche:2002ep,Zwanziger:2002ia,
Fischer:2002hna,Alkofer:2004it,Fischer:2006vf,Fischer:2007pf,newpaper} 
and functional renormalization group equations \cite{Pawlowski:2005xe,Pawlowski:2003hq}. 
In these continuum studies the dressing function of 
the ghost propagator is divergent, whereas the gluon propagator is 
infrared finite or even vanishing. In terms of the dressing functions $G(p^2)$ 
and $Z(p^2)$ of the ghost and gluon propagators in Landau gauge 
\begin{equation}
D_G(p) = -\frac{G(p^2)}{p^2}\,, \hspace*{1cm} 
 D_{\mu \nu}(p) = \left(\delta_{\mu \nu} -
    \frac{p_\mu p_\nu}{p^2}\right) D(p^2) = \left(\delta_{\mu \nu} -
    \frac{p_\mu p_\nu}{p^2}\right)
  \frac{Z(p^2)}{p^2}\,,
 \label{props} 
\end{equation}
and in terms of a power-law expansion in the infrared the dressing functions are related
by 
\begin{equation}
Z(p^2) \sim (p^2)^{2\kappa-\frac{d}{2}+2}; \hspace*{2cm} 
G(p^2) \sim (p^2)^{-\kappa}  \label{typeI} 
\end{equation} 
with dimension $d$ and positive and potentially irrational exponent $\kappa$.
These power laws are part of an all-order analytical analysis of both
the whole tower of DSEs and FRGs in the infrared \cite{Alkofer:2004it,Fischer:2006vf}.
They agree with a set of conditions formulated within a framework for confinement 
of covariantly gauge fixed Yang-Mills theory set up by Kugo and Ojima \cite{Kugo:1979gm}. 

The Kugo-Ojima scenario rests on well-defined charges related to unbroken global 
gauge symmetries. In particular it assumes global BRST symmetry. The related 
well defined charge operator has been used to identify the positive 
definite space $\cH_{phys}$ of physical states within the total 
state space $\cV$ of QCD. An unbroken global gauge symmetry 
is then crucial to show that the states in $\cH_{phys}$
contributing to the physical S-matrix of QCD are indeed
colorless. They also argued that this setup guarantees the 
disappearance of the 'behind-the-moon' problem, i.e. a 
colorless bound state with colored constituents cannot be 
delocalized into colored lumps \cite{Kugo:1979gm}. This then implements
the confining phase of Yang-Mills theory. In Landau gauge a direct
consequence of the well defined global color charge is the infrared
enhancement of the ghost dressing function $G(p^2)$ \cite{Kugo:1979gm}. Such a behavior
is obtained in eqs.(\ref{typeI}) if and only if $\kappa>0$. 

In functional methods this enhancement can be implemented as an
infrared renormalization condition for the ghost dressing function.
This condition leads to a unique
\cite{Fischer:2006vf} (scaling) solution of the whole tower of
functional equations for the one-particle irreducible Green's
functions of Yang-Mills theory. In turn, given the Kugo-Ojima scenario 
an infrared divergent ghost implicitly defines the unique
gauge fixing with well-defined global BRST-charges \cite{newpaper}.  

In lattice calculations, however, the behavior (\ref{typeI}) with $\kappa>0$ is 
notoriously difficult to obtain. In the two dimensional theory (\ref{typeI}) is 
nicely satisfied \cite{Maas:2007uv}. In three dimensions first hints of (\ref{typeI})
have been found in a formulation with an improved (evolutionary) gauge fixing algorithm
\cite{Maas:2008ri}, whereas in four dimensions one obtains (\ref{typeI}) in the strong 
coupling limit $\beta \rightarrow 0$ for not too small momenta \cite{lvs}. In general,
however, lattice calculations return the different behavior
\begin{equation}
Z(p^2) \sim (p^2)^{1/2}; \hspace*{2cm} 
G(p^2) \sim (p^2)^{0}  \label{typeII} 
\end{equation} 
even for very large lattices \cite{Cucchieri:2008fc}. This limit for 
$p^2 \rightarrow 0$, however, corresponds to a finite ghost dressing function
and is therefore not in agreement with the Kugo-Ojima scenario. Within functional methods
also this 'decoupling' type of solution can be implemented by suitable boundary conditions
in the infrared \cite{Lerche:2002ep,newpaper,Boucaud:2008ji,Aguilar:2008xm,Dudal:2008sp}. 
Up to logarithms, eqs.(\ref{typeI}) and (\ref{typeII}) 
completely exhaust the possible infrared solutions of the functional equations of 
Yang-Mills theory.

Given confinement, an infrared solution with finite ghost at zero momentum
(termed 'decoupling' below) implies broken global gauge and BRST
symmetries \cite{lvs,newpaper}. Indeed, all known BRST-quantizations that are compatible
with an infrared finite ghost even break off-shell BRST, see e.g.
\cite{Dudal:2008sp} and references therein. The only possibility for the 
decoupling solution to coexist with a globally well-defined BRST charge is 
in a Higgs phase, where the breaking of global color symmetry implies 
the existence of super-selection sectors. Certainly, this is not what is seen
in lattice simulations of QCD and therefore one may conclude that BRST-symmetry
is indeed broken on the lattice \cite{Neuberger:1986xz}. Regarding global symmetries, 
the status of the decoupling solutions is therefore clearly different from the 
scaling solution: whereas scaling agrees with well-defined BRST and global color charges
decoupling does not. Note, however, that both scaling and decoupling agree with 
the confinement criterion developed in \cite{Braun:2007bx}: both lead to a confining, 
nonperturbative Polyakov loop potential. Furthemore, in both cases
the gluon propagator exhibits positivity violation \cite{newpaper}.

\section{The ghost and gluon dressing functions}

\begin{figure}[t]
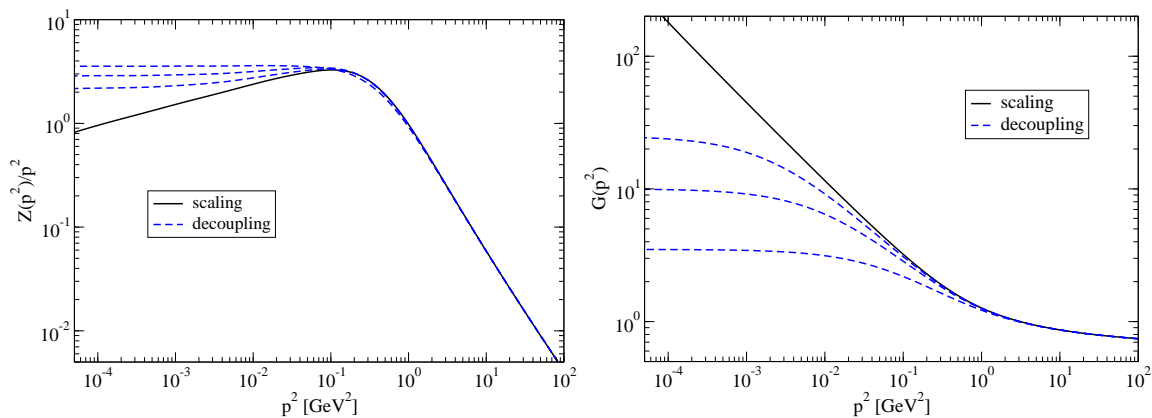

  \includegraphics[width=0.5\columnwidth]{newglue4.eps}
  \includegraphics[width=0.5\columnwidth]{newghost3.eps}
  \caption{Numerical solutions for the gluon propagator $D(p^2)=Z(p^2)/p^2$ 
    and the ghost dressing function $G(p^2)$ 
    with different boundary conditions $G(0)$.}
\label{res}
\end{figure}

As already mentioned above the functional continuum equations, DSEs and FRGs, 
can display both types of solutions, scaling (\ref{typeI}) and decoupling 
(\ref{typeII}). These are distinguished by a boundary condition for the ghost 
dressing function at zero momentum, $G(0)$. In \cite{newpaper} we demonstrated 
this behavior using a truncation scheme for the DSEs developed to guarantee the
transversality and multiplicative renormalizability of the gluon DSE. As an 
alternative we also employed a truncation for the corresponding equations in the
FRG-framework, which has been developed to minimize truncation artefacts in the 
mid-momentum region. 

Our numerical solutions for the ghost and gluon dressing functions are
shown in Fig.~\ref{res}. The boundary condition $G(0)=\infty$ results in 
the scaling solution, eq.~(\ref{typeI}), with a diverging ghost dressing 
function in the infrared and an infrared vanishing gluon propagator. The
corresponding critical exponent $\kappa$ in eq.~(\ref{typeI}) is given by 
$\kappa=\kappa_C=(93-\sqrt{1201})/98 \approx 0.595353$ \cite{Lerche:2002ep}. 
A finite value $G(0)=const.$, however, produces a continuous set of decoupling 
solutions with an infrared finite ghost dressing function. The corresponding 
gluon propagator is massive in the sense that 
$D(0)=\lim_{p^2 \rightarrow 0} Z(p^2)/p^2 = const.$ 
for decoupling. In the ultraviolet momentum region, both types of solutions 
are almost identical, as expected. 

\begin{figure}[t]
\centerline{\includegraphics[width=0.7\columnwidth]{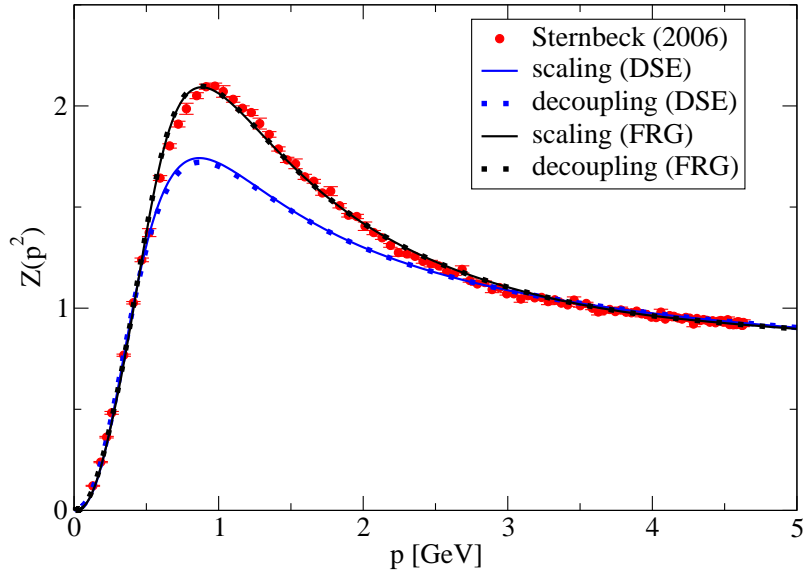}
}
  \caption{Both type of solutions compared to lattice results in 
  minimal Landau gauge from \cite{sternbeck06}.}
\label{latt}
\end{figure}

Finally we wish to emphasize that the question of scaling vs. decoupling only concerns 
global properties of the theory as the (non-)conservation of charges. The behavior
(\ref{typeI}) or (\ref{typeII}) sets in at scales $p^2 \ll \Lambda_{QCD}^2$. In
contradistinction all dynamics of the theory takes part on scales around or larger
than $\Lambda_{QCD}$. Certainly, from a phenomenological point of view the behavior 
of the ghost and gluon dressing function at scales $p^2 \ge \Lambda_{QCD}^2$ is 
much more relevant than the behavior in the deep infrared. In Fig.~\ref{latt} we 
compare the solutions from functional equations with the lattice results of 
ref.~\cite{sternbeck06} for the gluon dressing function. It is very satisfactory 
that our numerical solution of the functional renormalization group equations 
almost pointwise matches the corresponding lattice results in the phenomenologically
important mid-momentum region.

\vspace*{1cm}
{\bf Acknowledgments:}
We thank Reinhard Alkofer, Jens Braun, Jeff Greensite, Joannis Papavassiliou, Olivier
Pene, Kai Schwenzer and Lorenz von Smekal for useful discussions.
A.~M. was supported by the FWF under grant number P20330, C.~F by the
Helmholtz-University Young Investigator Grant number VH-NG-332
and J.~M.~P. by the ExtreMe Matter Institute (EMMI).\\[1mm]

\end{document}